\documentclass[letterpaper,twocolumn,flushbottom,prb,showpacs,floatfix,amsfonts,amssymb,amsmath,superscriptaddress,citeautoscript,balancelastpage]{revtex4-1}
\usepackage{graphicx}
\usepackage{color}
\usepackage[lofdepth,lotdepth,caption=false]{subfig}
\usepackage{amsmath}
\usepackage[pdfstartview=FitH]{hyperref}
\usepackage[usenames,dvipsnames,svgnames,table]{xcolor}
\usepackage{placeins}
\usepackage{comment}
\usepackage{txfonts} 
\usepackage[cal=cm]{mathalfa} 
\usepackage{cleveref} 
\crefname{equation}{Eq.}{Eqs.}
\crefname{figure}{Fig.}{Figs.}
\newsavebox{\bigleftbox}
\hypersetup{
  colorlinks,
  citecolor=Blue,
  linkcolor=Blue,
  urlcolor=Blue
}

\usepackage{pdfpages}
\usepackage{bookmark}

\graphicspath{{FIGS/}}

\begin{document}

\title{Unfolding spinor wave functions and expectation values of general operators: \\ Introducing the unfolding-density operator}

\author{Paulo V. C. Medeiros}
\email{paume@ifm.liu.se}
\email{paulovcmedeiros@gmail.com}
\affiliation{Department of Physics, Chemistry and Biology, IFM, Link\"oping University, 58183 Link\"oping, Sweden}

\author{Stepan S. Tsirkin}
\affiliation{Donostia International Physics Center (DIPC),  20018 San Sebasti\'an/Donostia, Basque Country, Spain}
\affiliation{Tomsk State University, 634050, Tomsk, Russia}
\affiliation{Saint Petersburg State University, Saint Petersburg 198504, Russia}

\author{Sven Stafstr{\"o}m}
\affiliation{Department of Physics, Chemistry and Biology, IFM, Link\"oping University, 58183 Link\"oping, Sweden}

\author{Jonas Bj\"ork}
\email{jonas.bjork@liu.se}
\affiliation{Department of Physics, Chemistry and Biology, IFM, Link\"oping University, 58183 Link\"oping, Sweden}

\date[]{Received 18 September 2014; published as a Rapid Communication in Phys. Rev. B on 26 January 2015}

\begin{abstract}
We show that the spectral weights $W_{m\vec K}(\vec k)$ used for the unfolding of two-component spinor eigenstates $\left| {\psi _{m\vec K}^\mathrm{SC}} \right\rangle = \left| \alpha \right\rangle \left| {\psi _{m\vec{K}}^\mathrm{SC, \alpha}} \right\rangle + \left| \beta \right\rangle \left| {\psi _{m\vec{K}}^\mathrm{SC, \beta}} \right\rangle$ can be decomposed as the sum of the partial spectral weights $W_{m\vec{K}}^{\mu}(\vec k)$ calculated for each component $\mu = \alpha, \beta$ independently, effortlessly turning a possibly complicated problem involving two coupled quantities into two independent problems of easy solution. Furthermore, we define the \emph{unfolding-density operator} $\hat{\rho}_{\vec{K}}(\vec{k}; \, \varepsilon)$, which unfolds the primitive cell expectation values $\varphi^{pc}(\vec{k}; \varepsilon)$ of any arbitrary operator $\mathbf{\hat\varphi}$ according to $\varphi^{pc}(\vec{k}; \varepsilon) = \mathit{Tr}\left(\hat{\rho}_{\vec{K}}(\vec{k}; \, \varepsilon)\,\,\hat{\varphi}\right)$. As a proof of concept, we apply the method to obtain the unfolded band structures, as well as the expectation values of the Pauli spin matrices, for prototypical physical systems described by two-component spinor eigenfunctions.
\end{abstract}

\maketitle

Modern electronic structure calculations, aided by the ever growing increase in computer power, aim more and more at tackling realistic problems. This is often done by means of supercell (SC) modeling, i.e., the use of a typically large unit cell whose lattice vectors $\vec{A}_{i}$ relate to the lattice vectors $\vec{a}_{j}$ of a given reference primitive cell (PC) as $\vec{A}_{i} = \sum_{j=1}^3N_{ij}\vec{a}_{j}$, with integers $N_{ij}$. In the ideal case, a SC is a perfect repetition of a given reference PC, meaning that not only the Bravais lattice, but also the positions of the atoms in the basis can be mapped from the SC to the PC. In practice, however, the effect of having defects, impurities and other types of perturbations is the very object of investigation, and thus the perfect mapping of the atomic positions is no longer possible. 

A procedure to unravel the PC Bloch character hidden in SC eigenstates is commonly referred to as \emph{unfolding}. Several unfolding approaches have been proposed \cite{Dargam1997, Wang1998, Boykin2005, Boykin2007, Ku2010, Zhang2011, Popescu2012, Allen2013, Medeiros2014, Tomic2014} and successfully applied to recover a PC representation of the band structure of systems described by means of both perfect and defective (often nearly perfect) SCs, greatly simplifying the analysis of the results and enabling direct comparisons with experimental measurements, such as angle-resolved photoemission spectroscopy (ARPES), often represented along the high-symmetry directions of the PC Brillouin zone (PCBZ). We use the expression ``nearly perfect SC'' to mean (i) SCs that deviate only slightly from a perfect repetition of a given reference PC, (ii) SCs consisting of a perfect repetition of the reference PC, combined with some weakly-interacting external agent(s), or (iii) a combination of (i) and (ii). The use of an unfolding methodology for a nearly perfect SC can be justified by considering the deviations from the ideal case as small perturbations \cite{Medeiros2014}. There are, nonetheless, scenarios in which unfolding can be justified however strong the influence of the presence of external agents might be, and those include, for instance, the assessment of how similar the eigenstates of a given system are to the eigenstates of its composing parts -- periodic systems themselves -- when not interacting with each other \cite{Allen2013}. For non-perfect cases, the unfolding yields an effective band structure (EBS) \cite{Popescu2012}.

Although such unfolding methodologies have successfully been used in conjunction with eigenvalue problems involving scalar wave functions, there has been little or no discussion so far, to the best of our knowledge, when it comes to spinor wave functions, despite the fact that the eigenstates of spin $1/2$ particles, such as electrons, are generally two-component spinors. This is particularly important, for instance, when the systems being modeled feature noncollinear magnetism or strong spin-orbit coupling \cite{DFT_spinors_noncollinear_mag1,DFT_spinors_non_collinear_mag2,VASP_noncollinear_mag}. Another important overlooked issue is the problem not only of unfolding the eigenvalues of the crystal Hamiltonian, but the more general one of unfolding the expectation values of any given operator, such as, for instance, the Pauli spin matrices. This is important for the study of , e.g., the spin polarization of graphene's $\pi$ bands induced by a heavy metal substrate \cite{Marchenko,Ru0001,Re0001,Ir111,Ir111_2}, as well as in Rashba-type splitting of Shockley surface states on reconstructed surfaces \cite{Nuber,PhysRevB.88.195433} and in surface alloys \cite{PhysRevLett.98.186807,PhysRevB.75.195414,PhysRevB.80.035438,PhysRevB.88.085427}.

In this work, we extend the unfolding methodology for the case of two-component spinor wave functions, and define the \emph{unfolding-density operator} $\hat{\rho}_{\vec{K}}(\vec{k}; \, \varepsilon)$, which unfolds
the PC expectation values $\varphi^{pc}(\vec{k}; \varepsilon)$ of any arbitrary operator $\hat{\varphi}$ according to ${\varphi^{pc}(\vec{k}; \varepsilon) = \mathit{Tr}\left(\hat{\rho}_{\vec{K}}(\vec{k}; \, \varepsilon)\,\,\hat{\varphi}\right)}$. To illustrate the applicability of the method, we perform some benchmark calculations on physically relevant model systems. 
%

In the following, $\{\vec{G}_{pcbz \leftarrow SCBZ}\}$ denotes the set of the ${\mathcal{N} \equiv {\Omega_\mathrm{pcbz}}/{\Omega_\mathrm{SCBZ}}}$ distinct SC reciprocal lattice (SCRL) translation vectors $\vec{G}_{i}$ that generate the PCBZ from the SC Brillouin zone (SCBZ), and $\{\vec{r}_{pc \rightarrow SC}\}$ is the set of the $\mathcal{N}$ distinct PC translation vectors $\vec{r}_{i}$ that generate the SC from the PC \cite{Allen2013}. The symbols ${\Omega _\mathrm{pcbz}}$ and ${\Omega _\mathrm{SCBZ}}$ represent, respectively, the volumes of the PCBZ and SCBZ. For every wave vector $\vec K$ of the SCBZ, there are thus $\mathcal{N}$ wave vectors $\vec k_i$ of the PCBZ obeying the geometric unfolding relation
\begin{equation}\label{unfolding_relation}
	{\vec k_i} = \vec K + \vec G_{i};\,\,\,\vec G_{i} \in \{\vec{G}_{pcbz \leftarrow SCBZ}\}.
\end{equation}
The unfolding theorem of Allen \emph{et al.} \cite{Allen2013} states that \emph{any function} $\Psi_{\vec{K}}(\vec{r})$ possessing the Bloch symmetry of the SC can be uniquely decomposed into a sum of partial functions ${\psi_{\vec{K} + \vec{G_{i}}}(\vec{r}) \equiv \hat{P}(\vec{K} \rightarrow \vec{K} + \vec{G_{i}}) \Psi_{\vec{K}}(\vec{r})}$, for every ${\vec G_{i} \in \{\vec{G}_{pcbz \leftarrow SCBZ}\}}$, each $\psi_{\vec{K} + \vec{G_{i}}}(\vec{r})$ satisfying ${\hat{T}(\vec{r}_{i}) \psi_{\vec{K} + \vec{G_{i}}}(\vec{r}) = e^{i(\vec{K} + \vec{G_{i}}) \cdot \vec{r}_{i}} \psi_{\vec{K} + \vec{G_{i}}}(\vec{r})}$, where $\hat{T}(\vec{r_{i}})$ denotes a translation by the PCRL vector $\vec{r_{i}}$. The projectors are given by \cite{Allen2013}
\begin{equation}\label{projectors_def}
\hat{P}(\vec{K} \rightarrow \vec{K} + \vec{G_{i}}) = \frac{1}{\mathcal{N}}\sum\limits_{\vec{r}_{j} \in \{\vec{r}_{pc \rightarrow SC}\}} \hat{T}(\vec{r_{j}})e^{-i(\vec{K} + \vec{G_{i}}) \cdot \vec{r_{j}}}.
\end{equation}
If $\Psi_{\vec{K}}(\vec{r})$ is normalized to unity, the norm of the partial function $\psi_{\vec{K} + \vec{G_{i}}}(\vec{r})$ can be used as a \textit{spectral weight} to assess the amount of PC Bloch character $\vec{k}_{i} = \vec{K} + \vec{G_{i}}$ hidden in $\Psi_{\vec{K}}(\vec{r})$. In particular, if $\left| {\psi _{m\vec K}^\mathrm{SC}} \right\rangle$ is an eigenstate of the Hamiltonian in the SC representation and $\vec{k}_{i}$ is a PCBZ wavevector related to ${\vec K}$ through \cref{unfolding_relation}, then the spectral weight ${W_{m\vec K}}(\vec{k}_{i})$ reads:
\begin{equation}\label{spec_weight_def_allen}
W_{m\vec K}(\vec{k}_{i}) \equiv  
\left\langle {\psi_{m\vec K}^\mathrm{SC}} | \hat{P}(\vec{K} \rightarrow \vec{k}_{i}) | {\psi _{m\vec K}^\mathrm{SC}} \right\rangle,
\end{equation}
where we have used ${\hat{P}^{2}(\vec{K} \rightarrow \vec{k}_{i}) = \hat{P}(\vec{K} \rightarrow \vec{k}_{i})}$.

Consider now the normalized two-component spinor eigenstates
\begin{equation}\label{definition_spinor_eigenstates}
\left| {\psi _{m\vec K}^\mathrm{SC}} \right\rangle = 
\left| \alpha \right\rangle 
\left| {\psi _{m\vec{K}}^\mathrm{SC, \alpha}} \right\rangle + 
\left| \beta \right\rangle 
\left| {\psi _{m\vec{K}}^\mathrm{SC, \beta}} \right\rangle
\end{equation}
satisfying the eigenvalue equation
\begin{equation}\label{eigenvalue_problem}
\hat{H}\left| {\psi _{m\vec K}^\mathrm{SC}} \right\rangle = \varepsilon_{m}(\vec{K}) \left| {\psi _{m\vec K}^\mathrm{SC}} \right\rangle
\end{equation}
for some crystal Hamiltonian $\hat{H}$. The ket spinors $\left| \alpha \right\rangle$ and $\left| \beta \right\rangle$ are the two eigenvectors of the Pauli spin matrix $\hat{\sigma}_{z}$:
\begin{equation}\label{definition_spinor_basis}
\left| \alpha \right\rangle = 
\left(\begin{array}{c}
       1 \\
       0  
       \end{array} 
\right);
\,\,
\left| \beta \right\rangle = 
\left(\begin{array}{c}
       0 \\
       1  
       \end{array} 
\right).
\end{equation}
For every $\vec{K}$, $\left\{ \left| {\psi _{m_{i}\vec K}^\mathrm{SC}} \right\rangle \right\}$ is a complete set orthonormal eigenfunctions of $\hat{H}$ with respect to the inner product
\begin{equation}\label{inner_prod_spinor_def}
\left\langle \varmathbb{F} | \varmathbb{G} \right\rangle = \varmathbb{F}^{\dagger} \varmathbb{G} = 
\left(\begin{array}{c c}
       F^{\alpha *} F^{\beta *}  
       \end{array} 
\right)
\left(\begin{array}{c}
       G^{\alpha} \\ 
       G^{\beta}  
       \end{array} 
\right)
= F^{\alpha *}G^{\alpha} + F^{\beta *}G^{\beta}.
\end{equation}
The unfolding theorem allows us to promptly arrive to a very useful result: Despite the fact that $\hat{H}$ generally couples the two components of $\left| {\psi _{m\vec K}^\mathrm{SC}} \right\rangle$, the spectral weights $W_{m\vec K}(\vec{k}_{i})$ can always be decomposed as
\begin{equation}\label{spec_weights_as_sum_of_partials}
W_{m\vec K}(\vec{k}_{i}) = W_{m\vec K}^{\alpha}(\vec{k}_{i}) + W_{m\vec K}^{\beta}(\vec{k}_{i}),
\end{equation}
where the \emph{partial spectral weights} $W_{m\vec K}^{\mu}(\vec{k}_{i})$ are defined as:
\begin{equation}\label{partial_spec_weight}
\begin{split}
W_{m\vec K}^{\mu}(\vec{k}_{i}) &\equiv 
\left\langle {\psi_{m\vec K}^\mathrm{SC, \mu}} | \hat{P}^{2}(\vec{K} \rightarrow \vec{k}_{i}) | {\psi _{m\vec K}^\mathrm{SC, \mu}} \right\rangle; \,\,\,\mu = \alpha, \beta.
\end{split}
\end{equation}
The reason is that the components of the spinor wave function $\left| {\psi _{m\vec K}^\mathrm{SC}} \right\rangle$ are not mixed when $\left| {\psi _{m\vec K}^\mathrm{SC}} \right\rangle$ is acted upon by the projectors $\hat{P}(\vec{K} \rightarrow \vec{k}_{i})$:
\begin{equation}\label{action_projector_spinor_state}
\hat{P}(\vec{K} \rightarrow \vec{k}_{i})
\left| {\psi _{m\vec K}^\mathrm{SC}} \right\rangle = 
\sum\limits_{\mu = \alpha, \beta}
\left| \mu \right\rangle
\left[ 
\hat{P}(\vec{K} \rightarrow \vec{k}_{i})
\left| {\psi _{m\vec{K}}^\mathrm{SC, \mu}} \right\rangle
\right].
\end{equation}
\cref{spec_weights_as_sum_of_partials} holds \emph{regardless of the basis set} used to represent $\left| {\psi _{m\vec K}^\mathrm{SC}} \right\rangle$. In fact, such a decomposition is reminiscent of the orbital decomposition presented in Ref. [\!\!\citenum{Ku2010}]. At no extra cost, \cref{spec_weights_as_sum_of_partials} turns the original problem, involving two possibly coupled quantities, into two completely independent problems. With this result, for instance, we straightforwardly generalize the expression for the number $N(\vec{k}; \varepsilon) = \lim_{\delta \varepsilon \rightarrow 0^{+}} \delta N(\vec{k}; \varepsilon)$ of unfolded PC bands crossing the point $(\vec{k}; \varepsilon)$ \cite{Medeiros2014} as 
\begin{equation}\label{unfolded_delta_Ns_spinors}
N(\vec{k}; \varepsilon) =
\sum\limits_{m} 
\sum\limits_{\mu = \alpha, \beta} W_{m\vec K}^{\mu}(\vec k)
\lim_{\delta \varepsilon \rightarrow 0^{+}}
\int\limits_{\varepsilon - \delta \varepsilon / 2}^{\varepsilon + \delta \varepsilon / 2}
\delta \left( {\varepsilon' - {\varepsilon_m}(\vec K)}\right)d\varepsilon'.
\end{equation}
%
%
%

We will now address a different problem, stated as follows: Suppose that $N(\vec{k_i} = \vec{K} + \vec{G_{i}}; \varepsilon) \ne 0$, i.e., that there is at least one PC band with energy $\varepsilon$ at the PCBZ wave vector $\vec{k_i}$. Given a general operator $\mathbf{\hat\varphi}$, and a complete set of SC eigenstates $\left| \psi _{m\vec K}^\mathrm{SC} \right\rangle$, how can one calculate the expectation value
\begin{equation}\label{defining_op_unfolding_problem}
\varphi^{pc}(\vec{k_{i}}; \varepsilon) \equiv 
\frac{1}{N(\vec{k_i}; \varepsilon)}
\sum_{\substack{n \\ \varepsilon_{n}(\vec{k_{i}}) = \varepsilon}}
\left\langle \psi_{n\vec{k}_{i}}^\mathrm{pc} \left| \mathbf{\hat\varphi} \right| \psi _{n\vec{k}_{i}}^\mathrm{pc} \right\rangle
\end{equation}
without explicitly calculating the PC eigenstates $\left| \psi _{n\vec{k}_{i}}^\mathrm{pc} \right\rangle$? We anticipate that $\varphi^{pc}(\vec{k}_{i}; \varepsilon)$ can be expressed as
\begin{equation}\label{unfolded_op_eigenvalues}
\varphi^{pc}(\vec{k}_{i}; \varepsilon) =
\mathit{Tr}\left(
\hat{\rho}_{\vec{K}}(\vec{k}_{i}; \, \varepsilon)
\,\,
\hat{\varphi}
\right),
\end{equation}
where $\hat{\rho}_{\vec{K}}(\vec{k}_{i}; \, \varepsilon)$ is completely defined by the geometric relations between the PC and SC lattice vectors. We refer to $\hat{\rho}_{\vec{K}}(\vec{k}_{i}; \, \varepsilon)$ as the \emph{unfolding-density operator}.

To find $\hat{\rho}_{\vec{K}}(\vec{k}_{i}; \, \varepsilon)$, we start by inserting the identity operator ${\openone = \sum_{m} 
\left| \psi _{m\vec K}^\mathrm{SC} \right\rangle
\left\langle \psi _{m\vec K}^\mathrm{SC} \right|
}$,
twice, in the right-hand side of \cref{defining_op_unfolding_problem}. After some rearrangement, this leads to 
\begin{equation}\label{op_unfolding_problem2}
\varphi^{pc}(\vec{k}_{i}; \varepsilon) =
\sum_{m' m} 
\frac{\varphi_{m' m}^{SC}(\vec{K})}{N(\vec{k_i}; \varepsilon)}
\left\langle \psi _{m\vec K}^\mathrm{SC} \right|
\left[
\sum_{\substack{n \\ \varepsilon_{n}(\vec{k}_{i}) = \varepsilon}}
\left| \psi _{n\vec{k}_{i}}^\mathrm{pc} \right\rangle
\left\langle \psi _{n\vec{k}_{i}}^\mathrm{pc} \right|
\right]
\left| \psi _{m'\vec K}^\mathrm{SC} \right\rangle,
\end{equation}
where $\varphi_{m' m}^{SC}(\vec{K}) \equiv \left\langle {\psi_{m'\vec K}^\mathrm{SC}} | \hat{\varphi}| {\psi _{m\vec K}^\mathrm{SC}} \right\rangle$. Since, for perfect SCs, ${\left\langle \psi_{n\vec{k}_{i}}^\mathrm{pc} | \psi _{m\vec K}^\mathrm{SC} \right\rangle = 0}$ if ${\varepsilon_{n}(\vec{k}_{i}) \ne \varepsilon_{m}(\vec{K})}$,
we can rewrite \cref{op_unfolding_problem2} as:
\begin{equation}\label{op_unfolding_problem3}
\varphi^{pc}(\vec{k}_{i}; \varepsilon) =
\sum_{\substack{m' m \\ \varepsilon_{m'}(\vec{K}) = \varepsilon \\ \varepsilon_{m}(\vec{K}) = \varepsilon}}
\frac{\varphi_{m' m}^{SC}(\vec{K})}{N(\vec{k_i}; \varepsilon)}
\left\langle \psi _{m\vec K}^\mathrm{SC} \right|
\left[
\sum_{n}
\left| \psi _{n\vec{k}_{i}}^\mathrm{pc} \right\rangle
\left\langle \psi _{n\vec{k}_{i}}^\mathrm{pc} \right|
\right]
\left| \psi _{m'\vec K}^\mathrm{SC} \right\rangle,
\end{equation}
where $n$ runs now over all PC bands. Notably,
\begin{equation}\label{equivalece_of_projs}
\sum_{n}
\left| \psi _{n\vec{k}_{i}}^\mathrm{pc} \right\rangle
\left\langle \psi _{n\vec{k}_{i}}^\mathrm{pc} \right|
= \hat{P}(\vec{K} \rightarrow \vec{k}_{i}),
\end{equation}
as the partial functions ${\hat{P}(\vec{K} \rightarrow \vec{k}_{i})\left| \psi _{m\vec K}^\mathrm{SC} \right\rangle}$ that decompose $\left| \psi _{m\vec K}^\mathrm{SC} \right\rangle$ according to the unfolding theorem belong to the subspace spanned by the eigenfunctions $\left| \psi _{n\vec{k}_{i}}^\mathrm{pc} \right\rangle$. Equation (\ref{op_unfolding_problem3}) then becomes:
\begin{equation}\label{unfolded_op_eigenvalues_discrete}
\varphi^{pc}(\vec{k}_{i}; \varepsilon) =
\sum_{\substack{m' m \\ \varepsilon_{m'}(\vec{K}) = \varepsilon \\ \varepsilon_{m}(\vec{K}) = \varepsilon}}
\varphi_{m' m}^{SC}(\vec{K})
\left\langle \psi _{m\vec K}^\mathrm{SC} \right|
\frac{\hat{P}(\vec{K} \rightarrow \vec{k}_{i})}{N(\vec{k_i}; \varepsilon)}
\left| \psi _{m'\vec K}^\mathrm{SC} \right\rangle.
\end{equation}
Let $\hat{\Lambda}_{\varepsilon}$ be an operator whose action on an arbitrary eigenstate $\left| \psi \right\rangle$ of $\hat{H}$ is to check whether  $\varepsilon_{\psi} \equiv {\left\langle \hat{H} \right\rangle}_{\psi}$ equals $\varepsilon$ or not. We define it in terms of its action on $\left| \psi \right\rangle$:
\begin{equation}\label{binary_op_def}
\hat{\Lambda}_{\varepsilon} 
\left| \psi \right\rangle =
\lambda_{\varepsilon, \varepsilon_{\psi}}
\left| \psi \right\rangle,
\end{equation}
where
\begin{equation}\label{eigenvalues_binary_op}
\lambda_{\varepsilon, \varepsilon_{\psi}} = 
\lim_{\delta \varepsilon \rightarrow 0^{+}}
\int\limits_{\varepsilon - \delta \varepsilon / 2}^{\varepsilon + \delta \varepsilon / 2}
\delta \left(\varepsilon' - \varepsilon_{\psi}\right)
d\varepsilon'.
\end{equation}
We can thus express \cref{unfolded_op_eigenvalues_discrete} as:
\begin{equation}\label{op_unfolding_problem4}
\varphi^{pc}(\vec{k}_{i}; \varepsilon) =
\sum_{m m'}
\left\langle \psi _{m\vec K}^\mathrm{SC} \right|
\frac{\hat{\Lambda}_{\varepsilon}
\hat{P}(\vec{K} \rightarrow \vec{k}_{i})
\hat{\Lambda}_{\varepsilon}}{N(\vec{k_i}; \varepsilon)}
\left| \psi _{m'\vec K}^\mathrm{SC} \right\rangle
\varphi_{m' m}^{SC}(\vec{K}),
\end{equation}
which is put into the form of \cref{unfolded_op_eigenvalues} by defining the unfolding-density operator $\hat{\rho}_{\vec{K}}(\vec{k}_{i}; \, \varepsilon)$ as
\begin{equation}\label{density_matrix}
\hat{\rho}_{\vec{K}}(\vec{k}_{i}; \, \varepsilon) \equiv
\frac{\hat{\Lambda}_{\varepsilon}
\hat{P}(\vec{K} \rightarrow \vec{k}_{i})
\hat{\Lambda}_{\varepsilon}}{N(\vec{k_i}; \varepsilon)}.
\end{equation}

The unfolding-density operator $\hat{\rho}_{\vec{K}}(\vec{k}_{i}; \, \varepsilon)$ has the properties of a mixed state density matrix. The condition ${\mathit{Tr}\left( \hat{\rho}_{\vec{K}}(\vec{k}_{i}; \, \varepsilon) \right) = 1}$ is verified by using ${\mathbf{\hat\varphi} = \openone}$ in \cref{unfolded_op_eigenvalues}, along with the definition of $N(\vec{k_i}; \varepsilon)$ [\cref{unfolded_delta_Ns_spinors}] and the fact that $\lambda_{\varepsilon, \varepsilon_{m}(\vec{K})}^{2} = \lambda_{\varepsilon, \varepsilon_{m}(\vec{K})}$. The Hermiticity of $\hat{\rho}_{\vec{K}}(\vec{k}_{i}; \, \varepsilon)$ is also immediate, as $N(\vec{k_i}; \varepsilon)$ is real and both $\hat{\Lambda}_{\varepsilon}$ and ${\hat{P}(\vec{K} \rightarrow \vec{k}_{i})}$ are Hermitian. Finally, $\hat{\rho}_{\vec{K}}(\vec{k}_{i}; \, \varepsilon) \ge 0$ follows by noticing that $W_{m\vec K}(\vec{k_{i}}), \, N(\vec{k_i}; \varepsilon), \, \lambda_{\varepsilon, \varepsilon_{m}(\vec{K})} \ge 0$.

To exemplify the use of the discussed formalism, we have obtained the unfolded band structures and unfolded expectation values of the Pauli vector $\vec{\sigma} \equiv \hat{\sigma}_{x}\hat{e}_{x} + \hat{\sigma}_{y}\hat{e}_{y} + \hat{\sigma}_{z}\hat{e}_{z}$ for some benchmark physical systems. As previously discussed, we justify the use of this unfolding methodology for nearly perfect SCs by considering the deviations from the ideal case as small perturbations. The methods have been implemented in BandUP \cite{Medeiros2014}, an open-source code freely available for download \cite{bandup_website}. Our DFT calculations, allowing for noncollinear magnetism and accounting for spin-orbit coupling effects, were performed using the VASP code \cite{VASP2, Perdew1996, VASP_noncollinear_mag}. These are typical cases where the formulation of the one-electron eigenvalue problem involves the use of two-component spinor eigenfunctions \cite{DFT_spinors_noncollinear_mag1,DFT_spinors_non_collinear_mag2,VASP_noncollinear_mag}. Specific computational details are given in the Supplemental Material (SM) \cite{SM}.

\begin{figure}[!t]
\includegraphics[width=.48\textwidth]{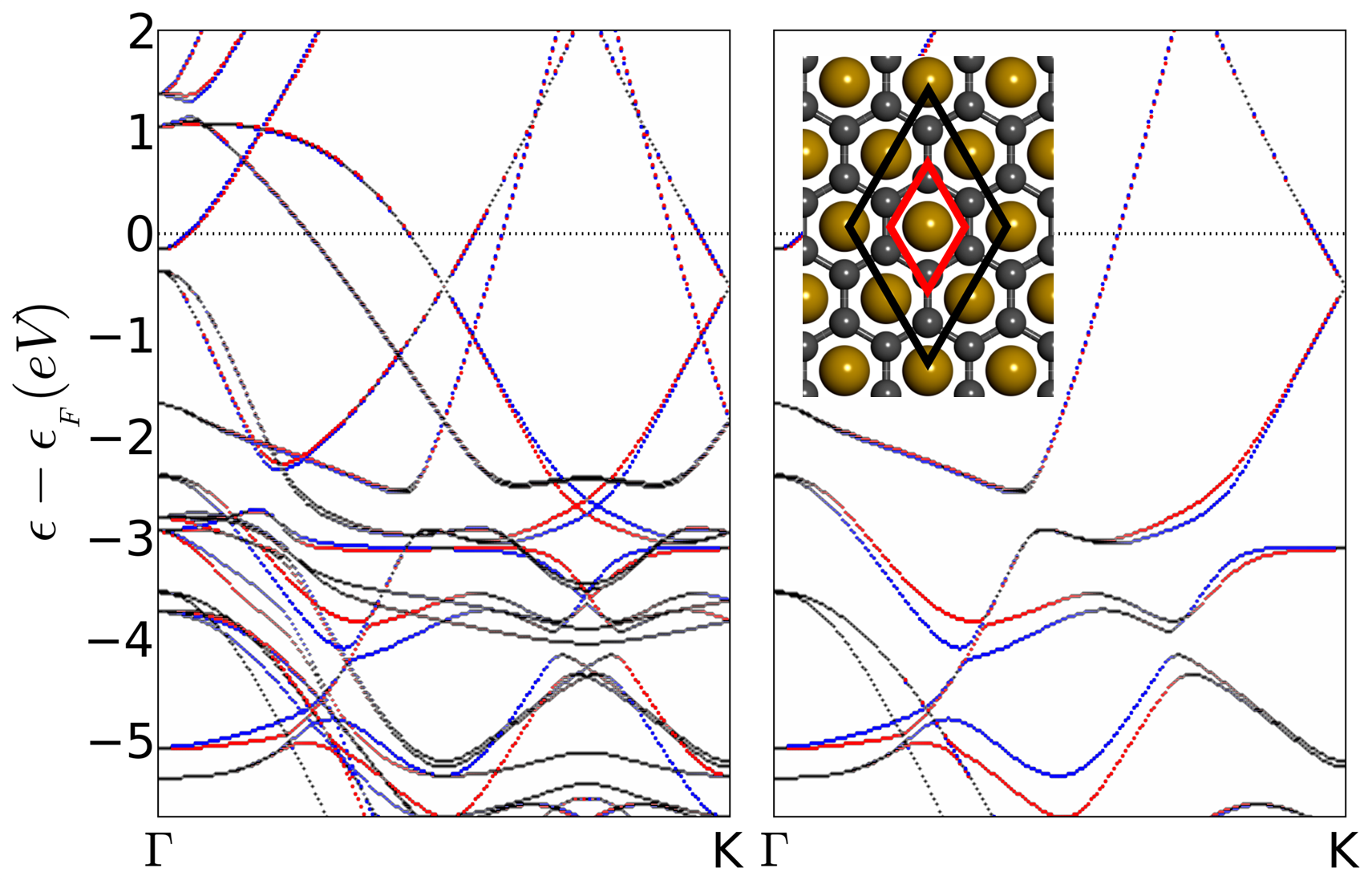}
\caption{
Graphene@Au: (a) Band structure and projections of the Pauli vector $\vec{\sigma} \equiv \hat{\sigma}_{x}\hat{e}_{x} + \hat{\sigma}_{y}\hat{e}_{y} + \hat{\sigma}_{z}\hat{e}_{z}$ for the $2\times2$ SC before (a) and after (b) unfolding onto the PC. The inset shows the $2 \times 2$ SC used (black rhombus), and a PC (red rhombus). In the curves, blue and red indicate opposite signs for the values of $\vec{\sigma}$ projected perpendicular to the PCBZ wavevectors, and black means a zero net value.
}
\label{fig:graphene_Au}
\end{figure}

As a first example, we consider an ideal case of a perfect SC. Figure \ref{fig:graphene_Au} shows the results of our simulations of a graphene layer with gold atoms attached on one side (one Au atom per graphene PC). Such a system has been used, for instance, as a model to understand the spin-orbit splitting in graphene, due to hybridization with gold, when graphene is adsorbed on an Ni(111) substrate with intercalated Au atoms \cite{Marchenko}. As systems with spin locked perpendicular to the momentum, such as Rashba-type spin-split surface states \cite{Datta1990} or surface states of three-dimensional topological insulators \cite{Hasan2010}, are considered promising for applications in spintronics, we calculated the spin projections perpendicular to the PCBZ wavevectors. Although spin-orbit splitting can be observed from the calculation involving the $2 \times 2$ SC, the use of the SC clearly complicates the analysis of the band structure and is prone to misleading interpretations. The unfolded band structure and eigenvalues of $\vec{\sigma}$ are also shown in \cref{fig:graphene_Au}. Since the SC is perfect, \cref{unfolded_delta_Ns_spinors,unfolded_op_eigenvalues} exactly recover the PC band/spin structure of the system, as reported in the SM \cite{SM}.  

\begin{figure}[!t]
\includegraphics[width=.36\textwidth]{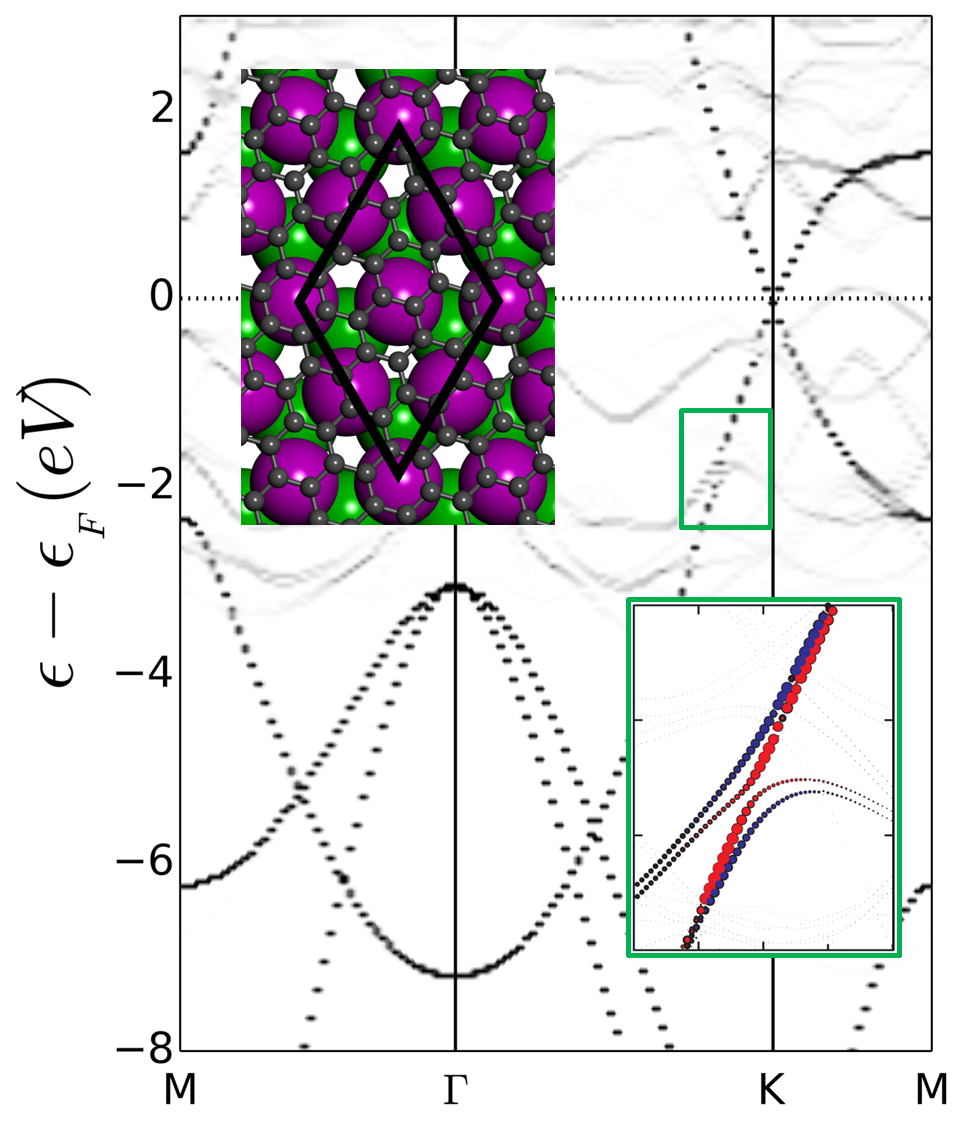}
\caption{
Graphene@Bi: EBS calculated along high-symmetry directions of graphene's PCBZ. The upper inset shows a top view of the atomic structure of the system, with the SC indicated by a black rhombus. The lower inset details the region delimited by the green rectangle. In the EBS inset, the colors are defined as in \cref{fig:graphene_Au} and the area of the spheres represents the magnitude of the projections.
}
\label{fig:graphene_Bi}
\end{figure}

Next, we consider the adsorption of graphene on a Bi(111) bilayer. Due to incommensurability between the two lattices \cite{Schiferl}, it is not possible to simulate graphene@Bi(111) with a single PC of graphene. Notably, this is often the case with epitaxially grown overlayers such as metal-organic interfaces \cite{Marks} and graphene on metal surfaces \cite{Ru0001,Re0001,Ir111,Ir111_2}. The in-plane lattice constant of Bi(111) is about 1.9 times greater than graphene's lattice constant, but a matching within 2\% is achieved for a $2 \times 2$ Bi(111) bilayer combined with $\sqrt{13} \times \sqrt{13}$ graphene, as shown in \cref{fig:graphene_Bi}. Graphene deviates only 0.02~\AA\,from being perfectly flat, and the graphene-Bi(111) equilibrium distance is of 3.4~\AA, incorporating van der Waals interactions in the calculations \cite{Grimme2006}. Since graphene interacts only weakly with bismuth, a picture of graphene's band structure in terms of its PCBZ is still useful. Figure \ref{fig:graphene_Bi} shows the EBS obtained for the system. While the calculated folded band structure \cite{SM} is practically unreadable, the signature of a quasi-freestanding graphene layer is clearly featured in the EBS. Strikingly, the effects of the interaction with the Bi substrate are directly revealed by unfolding the expectation values of $\vec{\sigma}$: In the regions of intersection between graphene and Bi bands, spin-dependent avoided-crossing effects appear, causing spin-splitting of the graphene bands (see \cref{fig:graphene_Bi}, inset).

\begin{figure}[!t]
\includegraphics[width=.35\textwidth]{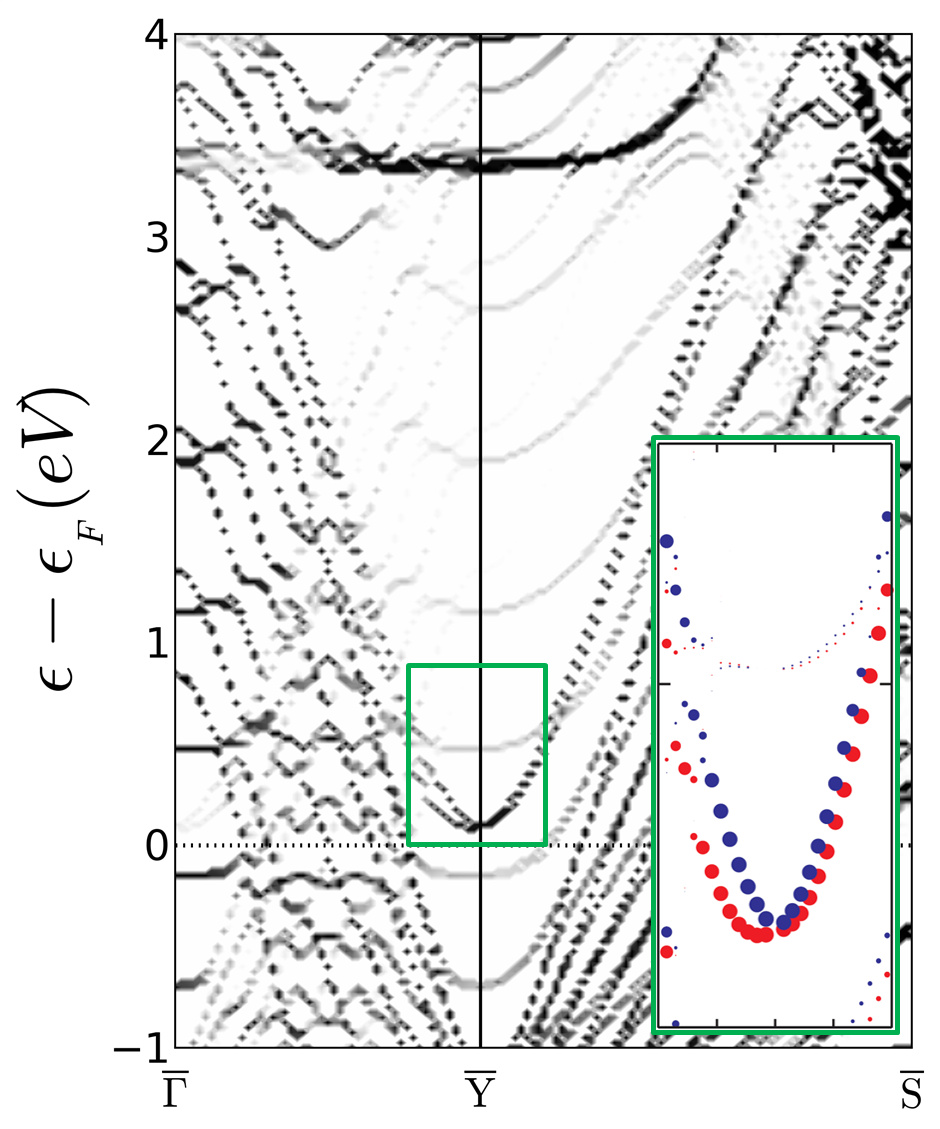}
\caption{
Au(110), $2 \times 1$ reconstructed surface: EBS along high-symmetry directions of the PCBZ of the non-reconstructed surface. The inset shows a zoom-in into the region delimited by the green rectangle. In the inset, the colors are defined as in \cref{fig:graphene_Au}, but the projections are now onto the perpendicular to $\vec{k} - \overline{Y}$. The area of the spheres represents the magnitude of the projections.
}
\label{fig:Au110}
\end{figure}

Our final example is the $2\times1$ reconstructed Au(110) surface \cite{Nuber}. Under reconstruction, the $\overline{Y}$ gap, containing two Shockley states (at -0.6 and +1.35 eV for the unreconstructed surface), folds into the $\overline{\Gamma}$ point, where the continuum of bulk states exists. However, by unfolding the band structure onto the PCBZ (\cref{fig:Au110}), we clarify that the lower surface state survives as a surface resonance around the $\overline{Y}$ point. The SC bulk states are also unfolded from the $\overline{\Gamma}$ point to the PCBZ $\overline{Y}$ point, but with very small $N(\vec{k}; \varepsilon)$/spectral weights, forming nothing but a weak background that introduces only little broadening to the surface resonance. Therefore, such resonance can, in practice, be considered as a surface state. Since the reconstruction pushes the surface state above the Fermi level, it is not detected by ARPES \cite{Nuber}. The same experiment, nevertheless, undoubtedly detects the energy gap at the $\overline{Y}$ point. The surface state has anisotropic Rashba-type spin splitting. The unfolding of the eigenvalues of $\vec{\sigma}$ enables the quantification of the splitting for the $\overline{Y\Gamma}$ and $\overline{YS}$ directions: $\Delta k=$ 0.055~\AA$^{-1}$ and 0.017~\AA$^{-1}$, respectively.

%

In conclusion, we have shown that the spectral weights for the unfolding of two-dimensional spinors can always be decomposed as the sum of partial spectral weights, one for each spinor component, transforming, at no extra cost, a problem of two possibly coupled quantities into two independent tractable problems. In a plane wave basis set, both the total and the partial spectral weights take the same form as the one for scalar wave functions (see SM \cite{SM}). We introduced the \emph{unfolding-density operator}, which unfolds the primitive cell expectation values for any given operator directly from a super cell calculation, extending the unfolding methodology to any $k$-space sensitive property. The applicability of the method was demonstrated for systems described in terms of two-component spinors, in particular to unfold expectation values of the Pauli spin matrices.

Given the general and basis-set independent character of our discussion, we believe that our work can be adapted to more complex cases without major complications. The development and implementation of methods to unfold band structures is a very active topic of research, which has already brought up many intriguing questions and answers. Indeed, by the time this work was being processed by the publisher, a related approach was used to unfold the Berry curvature using Wannier Functions \cite{Bianco2014}. Besides extending the scope of the discussion to the unfolding of other material properties, we anticipate that our results will motivate researchers to tackle other emerging problems. There is no doubt that, given the rapid recent developments in both theory and computational implementation, the unfolding methodologies being developed now will soon become common practices in the study of periodic materials.  
%
\clearpage
P.V.C.M, S.S. and J.B. acknowledge the Swedish Research Council (VR) for funding. S.S.T. acknowledges funding from the University of Basque Country UPV/EHU (GIC07-IT-756-13), the Departamento de Educaci\'{o}n del Gobierno Vasco and the Spanish Ministerio de Ciencia e Innovaci\'{o}n (FIS2010-19609-C02-01), the Tomsk State University Competitiveness Improvement Program, the Saint Petersburg State University (project 11.50.202.2015), and the Spanish Ministry of Economy and Competitiveness MINECO (FIS2013-48286-C2-1-P). Computer resources were allocated by the National Supercomputer Centre, Sweden, through SNAC and the MATTER consortium, as well as in the SKIF-Cyberia and CRYSTAL supercomputers at Tomsk State University. \textbf{NOTE:} This manuscript is a preprint of the paper \mbox{Phys. Rev. B \textbf{91} 041116(R), (2015)}. When referring to the work presented here, please cite the published PRB paper instead.

\bibliography{references}

\onecolumngrid
\newpage
\bookmark[page=6,level=0,view={FitH \calc{\paperheight-\topmargin-1in}}]{Supplemental Material}


\section*{Supplementary material}

# Supplemental Material for
# "Unfolding spinor wave functions and expectation values of general operators: Introducing the unfolding-density operator"

Paulo V. C. Medeiros,[1, *] Stepan S. Tsirkin,[2, 3, 4] Sven Stafström,[1] and Jonas Björk[1, †]

[1]*Department of Physics, Chemistry and Biology, IFM, Linköping University, 58183 Linköping, Sweden*
[2]*Donostia International Physics Center (DIPC), 20018 San Sebastián/Donostia, Basque Country, Spain*
[3]*Tomsk State University, 634050, Tomsk, Russia*
[4]*Saint Petersburg State University, Saint Petersburg 198504, Russia*

## COMPUTATIONAL DETAILS

Our benchmark calculations were carried out using density-functional theory with a plane-wave basis set expansion of the Kohn-Sham orbitals. We used PAW potentials [1, 2], with cutoffs of 400 eV for Au@Graphene and Graphene@Bi, and 250 eV for Au(110). The calculations were performed allowing for noncollinear magnetism and accounting for spin-orbit coupling effects [3]. We used the GGA/PBE approximation for Au@Graphene and Graphene@Bi, and LDA for the reconstructed Au(110) surface. Van der Waals interactions were included in the calculations involving Graphene@Bi [4]. In all self-consistent calculations, the Brillouin zones were sampled by using Monkhorst-Pack k-point grids. A gamma-centered scheme was employed in the simulations of Au@Graphene and Graphene@Bi, with grids of $16 \times 16 \times 1$ and $8 \times 8 \times 1$ special k-points for Au@Graphene (PC and SC), and $3 \times 3 \times 1$ for Bi@Graphene. For Au(110), a $12 \times 4 \times 1$ grid was used. The slab employed for the simulation of the reconstructed Au(110) surface had 21 layers, and the atoms in the 5 outermost layers were allowed to fully relax until the forces were all smaller than 0.01 eV/Å.

The self-consistent cycles were considered to be converged when the differences between the total energies, as well as between Kohn-Sham eigenvalues for the same orbitals, in two consecutive cycles, were smaller than $1 \times 10^{-6}$ eV (relaxations) and $1 \times 10^{-7}$ eV (band structures). The unfolding method discussed in the main text has been implemented in the BandUP code [5, 6]. The unfolding was performed using an energy grid with intervals of 1 meV for the numerical results and 25 meV for the figures.

## COMPLEMENTARY RESULTS

### Spectral Weights for Spinor Eigenstates: Plane-wave Basis Set

As a particular case of the decomposition of the spectral weights $W_{m\vec{K}}(\vec{k}_j)$ for spinor eigenstates [Eq. (8) of the main text], consider the plane-wave representation of $\left|\psi_{m\vec{K}}^{SC}\right\rangle$:

$$\left\langle \vec{r} | \psi_{m\vec{K}}^{SC} \right\rangle = \sum_{\vec{G} \in SCRL} \mathbb{C}_{m\vec{K}}^{SC}(\vec{G}) e^{i(\vec{K}+\vec{G}) \cdot \vec{r}}, \tag{S1}$$

where

$$\mathbb{C}_{m\vec{K}}^{SC}(\vec{G}) = \sum_{\mu=\alpha,\beta} |\mu\rangle C_{m\vec{K}}^{SC,\mu}(\vec{G}) \tag{S2}$$

and $C_{m\vec{K}}^{SC,\mu}(\vec{G})$ are generally complex. Similarly to the procedure adopted in Ref. [7], we use the mapping provided by the geometric unfolding relations [Eq. (1)] to rewrite Eq. (S1) as:

$$\left\langle \vec{r} | \psi_{m\vec{K}}^{SC} \right\rangle = \sum_{\substack{\vec{g} \in pcrl \\ \vec{G}_i \in \{\vec{G}_{pcbz \leftarrow SCBZ}\}}} \mathbb{C}_{m\vec{K}}^{SC}(\vec{g} + \vec{G}_i) e^{i(\vec{K}+\vec{g}+\vec{G}_i) \cdot \vec{r}}. \tag{S3}$$

It thus follows that

$$\hat{P}(\vec{K} \to \vec{K} + \vec{G}_j) \left\langle \vec{r} | \psi_{m\vec{K}}^{SC} \right\rangle = \sum_{\vec{g} \in pcrl} \mathbb{C}_{m\vec{K}}^{SC}(\vec{g} + \vec{G}_j) e^{i(\vec{K}+\vec{g}+\vec{G}_j) \cdot \vec{r}}, \tag{S4}$$

where we used $e^{i\vec{g}\cdot\vec{r}_l} = 1$ and also [8]

$$\frac{1}{\mathcal{N}} \sum_{\vec{r}_l \in \{\vec{r}_{pc \to SC}\}} e^{i(\vec{G}_i+\vec{G}_j)\cdot\vec{r}_l} = \delta_{i,j}; \quad \vec{G}_i, \vec{G}_j \in \{\vec{G}_{pcbz \leftarrow SCBZ}\}. \tag{S5}$$

Combining Eqs. (3), (7) and S4, making the substitution $\vec{k}_j = \vec{K} + \vec{G}_j$, and using $\hat{P}^2(\vec{K} \to \vec{k}) = \hat{P}(\vec{K} \to \vec{k})$ and $\left\langle e^{i\vec{g}\cdot\vec{r}} | e^{i\vec{g}'\cdot\vec{r}} \right\rangle = \delta_{\vec{g}\vec{g}'}$, we finally arrive at:

$$W_{m\vec{K}}(\vec{k}_j) = \sum_{\vec{g} \in \text{pcrl}} \left| \mathbb{C}^{SC}_{m\vec{K}}(\vec{g} + \vec{G}_j) \right|^2 = W^{\alpha}_{m\vec{K}}(\vec{k}) + W^{\beta}_{m\vec{K}}(\vec{k}), \tag{S6}$$

where

$$W^{\mu}_{m\vec{K}}(\vec{k}_j) = \sum_{\vec{g} \in \text{pcrl}} \left| C^{SC,\mu}_{m\vec{K}}(\vec{g} + \vec{G}_j) \right|^2; \quad \mu = \alpha, \beta. \tag{S7}$$

Eqs. (S6) and (S7) have the same form as the one derived in Ref. [7], except that, in Eq. (S6), the plane-wave coefficients are now spinor objects instead of scalar quantities, and, in Eq. (S7), the coefficients relate to the two components of the same spinor eigenfunction.

### Matrix Elements of the Unfolding-Density Operator in a Plane-Wave Basis Set

According to the definitions of the operators $\hat{\Lambda}_\varepsilon$ and $\hat{\rho}_{\vec{K}}(\vec{k}_j; \varepsilon)$ [Eqs. (18), (19) and (21)], the matrix elements of $\hat{\rho}_{\vec{K}}(\vec{k}_j; \varepsilon)$ in the $\left\{ \left| \psi^{SC}_{m\vec{K}} \right\rangle \right\}$ basis are:

$$\hat{\rho}_{m'm;\vec{K}}(\vec{k}_j; \varepsilon) = \frac{\lambda_{\varepsilon,\varepsilon_{m'}(\vec{K})} \lambda_{\varepsilon,\varepsilon_m(\vec{K})}}{N(\vec{k}_j; \varepsilon)} W_{m'm\vec{K}}(\vec{k}_j), \tag{S8}$$

where

$$W_{m'm\vec{K}}(\vec{k}_j) \equiv \left\langle \psi^{SC}_{m'\vec{K}} | \hat{P}(\vec{K} \to \vec{k}_j) | \psi^{SC}_{m\vec{K}} \right\rangle. \tag{S9}$$

The diagonal elements of $\left( W_{m'm\vec{K}}(\vec{k}_j) \right)$ are the spectral weights $W_{m\vec{K}}(\vec{k}_j)$. Combining Eqs. (11), (19) and S9 thus gives:

$$N(\vec{k}_i; \varepsilon) = \sum_m \lambda_{\varepsilon,\varepsilon_m(\vec{K})} W_{mm\vec{K}}(\vec{k}) \tag{S10}$$

Following the same steps as in the previous section, Eq. (S9) is expressed in a plane-wave basis set as

$$W_{m'm\vec{K}}(\vec{k}_j) = \sum_{\vec{g} \in \text{pcrl}} \mathbb{C}^{\dagger SC}_{m'\vec{K}}(\vec{g} + \vec{G}_j) \mathbb{C}^{SC}_{m\vec{K}}(\vec{g} + \vec{G}_j). \tag{S11}$$

Therefore, Eq. (S8) becomes

$$\hat{\rho}_{m'm;\vec{K}}(\vec{k}_j; \varepsilon) = \frac{\lambda_{\varepsilon,\varepsilon_{m'}(\vec{K})} \lambda_{\varepsilon,\varepsilon_m(\vec{K})}}{N(\vec{k}_i; \varepsilon)} \sum_{\vec{g} \in \text{pcrl}} \mathbb{C}^{\dagger SC}_{m'\vec{K}}(\vec{g} + \vec{G}_j) \mathbb{C}^{SC}_{m\vec{K}}(\vec{g} + \vec{G}_j), \tag{S12}$$

where

$$N(\vec{k}_i; \varepsilon) = \sum_{\substack{m \\ \vec{g} \in \text{pcrl}}} \lambda_{\varepsilon,\varepsilon_m(\vec{K})} \left| \mathbb{C}^{SC}_{m\vec{K}}(\vec{g} + \vec{G}_j) \right|^2. \tag{S13}$$

Naturally, Eqs. (S12) and (S13) are valid for both spinor and scalar wave functions.

**Band Structures and Spin Projections**

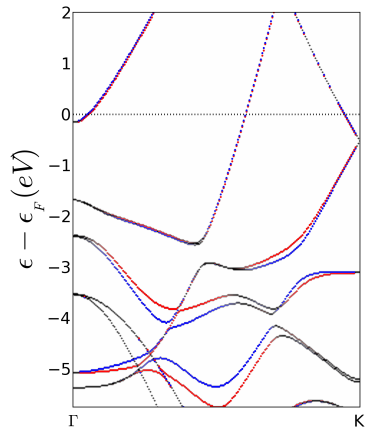

FIG. S1. Graphene@Au: Band structure and projections of the Pauli vector $\vec{\sigma} \equiv \hat{\sigma}_x \hat{e}_x + \hat{\sigma}_y \hat{e}_y + \hat{\sigma}_z \hat{e}_z$ calculated using the PC. Blue and red colors indicate opposite signs for the values of $\vec{\sigma}$ projected perpendicular to the PCBZ wavevectors, and black means a zero net value.

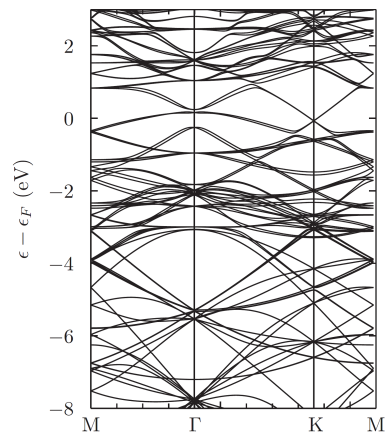

FIG. S2. Band structure of graphene@Bi. Notice that the calculation refers to the BZ of the system as a whole, and not to graphene's PCBZ.

Fig. S1 displays the band structure and projections of the Pauli vector for the graphene@Au system, calculated using the PC directly. As expected, this result is reproduced by performing unfolding from the $2 \times 2$ SC calculation, as discussed in the main text. Fig. S2 shows the folded band structure for graphene@Bi. The folding of the bands greatly hinders the analysis of the important aspects of the band structure of graphene, specially because the BZs of graphene and graphene@Bi are rotated with respect to each other.



\end{document}